\documentclass{aastex63}

\usepackage{bm}
\received{\today}
\revised{\today}
\accepted{\today}
\submitjournal{ApJL}

\shorttitle{Generation of Compressive Alf\'enic Pulses in Solar Wind Source Region}
\shortauthors{He et al.}

\graphicspath{{./}{figures/}}

\begin{document}
\title{Possible Generation Mechanism for Compressive Alfv\'enic Spikes as Observed by Parker Solar Probe}

\correspondingauthor{Jiansen He, Liping Yang}
\email{jshept@pku.edu.cn; lpyang@swl.ac.cn}

\author[0000-0001-8179-417X]{Jiansen He}
\affiliation{School of Earth and Space Sciences, Peking University, \\
Beijing, 100871, P. R. China}

\author[0000-0002-1541-6397]{Xingyu Zhu}
\affiliation{School of Earth and Space Sciences, Peking University, \\
Beijing, 100871, P. R. China}

\author{Liping Yang}
\affiliation{State Key Laboratory of Space Weather, National Space Science Center, Chinese Academy of Sciences, \\
Beijing, 100190, P. R. China}

\author{Chuanpeng Hou}
\affiliation{School of Earth and Space Sciences, Peking University, \\
Beijing, 100871, P. R. China}

\author[0000-0002-6300-6800]{Die Duan}
\affiliation{School of Earth and Space Sciences, Peking University, \\
Beijing, 100871, P. R. China}

\author{Lei Zhang}
\affiliation{Qian Xuesen Laboratory of Space Technology, \\
Beijing, 100190, P. R. China}

\author{Ying Wang}
\affiliation{School of Earth and Space Sciences, Peking University, \\
Beijing, 100871, P. R. China}

\begin{abstract}
The solar wind is found by Parker Solar Probe (PSP) to be abundant with Alfv\'enic velocity spikes and magnetic field kinks. Temperature enhancement is another remarkable feature associated with the Alfv\'enic spikes. How the prototype of these coincident phenomena is generated intermittently in the source region becomes a hot topic of wide concerns. Here we propose a new model introducing guide-field discontinuity into the interchange magnetic reconnection between open funnels and closed loops with different magnetic helicities. The modified interchange reconnection model not only can accelerate jet flows from the newly opening closed loop but also excite and launch Alfv\'enic wave pulses along the newly-reconnected and post-reconnected open flux tubes. We find that the modeling results can reproduce the following observational features: (1) Alfv\'en disturbance is pulsive in time and asymmetric in space; (2) Alfv\'enic pulse is compressive with temperature enhancement and density variation inside the pulse. We point out that three physical processes co-happening with Alfv\'en wave propagation can be responsible for the temperature enhancement: (a) convection of heated jet flow plasmas (decrease in density), (b) propagation of compressive slow-mode waves (increase in density), and (c) conduction of heat flux (weak change in density). We also suggest that the radial nonlinear evolution of the Alfv\'enic pulses should be taken into account to explain the formation of magnetic switchback geometry.
\end{abstract}
\keywords{}

\section{Introduction} \label{sec:intro}
Thanks to the observations with high time resolution and high optical sensitivity from state-of-the-art facilities, various dynamic phenomena and processes (e.g., magnetic reconnections, jet flows, and oscillatory waves) have been observed to be omnipresent in the multi-layers of solar atmosphere \citep{Shibata2007, Tomczyk2007, He2010, Tian2014, Shen2018}. In the past, magnetic reconnection and wave dissipation are viewed as two seemingly distinct mechanisms opposite to one another when dealing with the problems of coronal heating and solar wind origin \citep[e.g.,][]{Cranmer2010}. Magnetic reconnection rapidly converts magnetic energy to particle energy, causing emission flare in multiple wavebands and even triggering coronal mass ejections \citep{Priest2000}. Nano-jets are observed in closed loop system and suggested to be caused by slingshot effect of newly reconnected magnetic field lines with small angle of shear before reconnection \citep{Antolin2020}. Oscillatory wave signatures (e.g., oscillations in radiation intensity and Doppler velocities) have been probed to be propagating or standing in both open strands or closed loops throughout the solar atmosphere \citep{Wang2009}. Flares are found to be often accompanied by quasi-periodic pulsations (QPPs) in multi-band radiation intensities \citep{Nakariakov2010, Van2016, Mclaughlin2018}. The causal relation between reconnection and Alfv\'en waves was explored in 3D MHD simulation by Wal\'en test of the field-aligned current and parallel vorticity as emitting from the reconnection site \citep{Ma1995}. In the chromosphere, excitation of kink or Alfv\'enic waves as well as compressive slow-mode waves are also discovered to be associated with the formation of type II spicules, which are launched by magnetic reconnection \citep{He2009kink, He2009upward, Liu2014}.  

The MHD waves as generated in the solar atmosphere can leak their residual signals into the interplanetary space after reflection, dissipation, and mode conversion during their outward journey. Therefore, the solar wind turbulence looks wavelike and abundant with various waves. Outward Alfv\'en waves are prevalent in the fast streams, playing a vital role in accelerating and heating the plasmas in the fast streams \citep{Belcher1971, Tu1997, Li2004, Suzuki2005, Cranmer2007, Matsumoto2014, Yang2016}. A pronounced feature of compressive fluctuations is the multi-scale pressure balance: a balance of total pressure (consisting of thermal and magnetic pressures) between adjacent flux tubes, indicating the cascading process of quasi-perpendicular slow-mode waves \citep{Yao2011, Yang2017, Verscharen2019}. Quasi-parallel slow-mode waves are more likely to be Landau damped than their quasi-perpendicular counterparts of the same wavenumber. Co-existence of Alfv\'en waves and slow-mode waves are often observed in the solar wind streams, shaping the proton velocity distribution to form multi-beam populations \citep{He2015slow}.  Bi-directional Alfv\'en waves have been found in reconnection exhaust region due to parallel energization of protons and consequent trigger of firehose instability growth \citep{He2018}.

Large amplitude Alfv\'enic fluctuations, a phenomenon frequently observed in the solar wind \citep{Wang2012large}, often show the profiles of one-side pulses \citep{Gosling2009}. The interplanetary magnetic field lines as disturbed by the one-side Alfv\'enic pulses would display a spiral geometry intermittently distorted by sudden kinks rather than superposed with sinusoidal wave trains. A reversal of magnetic field polarity can happen when the fluctuation amplitude is large enough \citep{Yamauchi2004, Horbury2020}. The sudden polarity reversal of magnetic field components has been reported at various distances for several decades since the era of Helios mission, which observed the reversal events in the inner heliosphere at the distance as close as 0.3 AU from the Sun \citep{Horbury2018}. The reversal of magnetic field can also be called as deflection or switchback of magnetic field line. The associated sudden jumps in velocity are named as plasma jets or Alfv\'enic spikes/pulses \citep{Matteini2014}. The Alfv\'enic switchback as observed by PSP during its journey towards the perihelion again attracts wide attentions and sparks a hot debate \citep{Kasper2019, Bale2019, deWit2020, Krasnoselskikh2020}. A initially preset switchback configuration in a domain with periodic boundary conditions can survive in MHD simulation for several hundreds of Alfv\'en transit times, which may correspond to the propagation time from the solar corona to the perihelion of PSP \citep{Tenerani2020}. Interchange magnetic reconnection between closed and open field lines is argued to be responsible for the escape of switchback and for the global transportation of open magnetic flux \citep{Fisk2020}. The compressibility associated with the Alfv\'enic jumps of the switchback structures are theoretically argued to be a signature of fast-magnetosonic mode \citep{Zank2020}.

How is the Alfv\'enic switchback or its prototype generated in the solar atmosphere self-consistently? Why is the Alfv\'enic switchback usually associated with thermal and compressive fluctuations? What can we learn about the physical process taking place at the root of open magnetic field lines from the encounter of Alfv\'enic switchback in the inner heliosphere? We are trying to address these interrelated issues in this work.

\section{Model setup for Interchange Reconnection with Guide Field Discontinuity}
We adopt the resistive MHD equations in the Cartesian coordinates (x, y, z) with y being the vertical coordinate and x-z being the horizontal plane parallel to the solar surface. The numerical model is conducted in two dimensions (x-z) for the vectors of three components. The dimensionless form of the MHD equations for the physical variables (density $\rho$, momentum $\rho \bf{u}$, energy $e$, and magnetic field $\bf{B}$) read as followings:

  \begin{eqnarray}
\frac{\partial \rho}{\partial t}+\nabla \cdot \rho \mathbf{u} &=& 0, \\ 
\frac{\partial \rho \mathbf{u}}{\partial t}+\nabla \cdot\left[\rho \mathbf{u} \mathbf{u}+\mathbf{I}\left(p+\frac{1}{2} \mathbf{B}^{2}\right)-\mathbf{B B}\right] &=& \rho \mathbf{g}, \\ 
\frac{\partial e}{\partial t}+\nabla \cdot\left[\mathbf{u}\left(e+p+\frac{1}{2} \mathbf{B}^{2}\right)-(\mathbf{u} \cdot \mathbf{B}) \mathbf{B}\right] 
&=& \rho \mathbf{u} \cdot \mathbf{g}+\nabla \cdot(\mathbf{B} \times \eta \mathbf{j})-L_{r}+\nabla \cdot \mathbf{q}+H+C_{N}, \\ 
\frac{\partial \mathbf{B}}{\partial t}+\nabla \cdot(\mathbf{u B}-\mathbf{B u}) &=& \eta \nabla^{2} \mathbf{B},
 \end{eqnarray}
where 
  \begin{eqnarray}
	e=\frac{1}{2} \rho \mathbf{u}^{2}+\frac{p}{\gamma-1}+\frac{1}{2} \mathbf{B}^{2}, \quad \mathbf{j}=\nabla \times \mathbf{B}.
 \end{eqnarray}

The source terms in the energy equation are considered to account for the physical processes: work done by solar gravity ($\mathbf{g}$), effect of magnetic resistivity ($\eta$), radiation cooling ($-L_{\rm{r}}$), divergence of thermal conduction flux ($\nabla\cdot\mathbf{q}$), background heating ($H$), additional Newton cooling for chromosphere maintaince ($C_{\rm{N}}$). A splitting-based finite volume method (FVM) is applied for two different parts: the second-order Godunov-type central scheme for the fluid part; the constrained transport (CT) approach for the magnetic field part. The second-order Runge–Kutta method is implemented for the time advancement. The discrete grid is dynamically allocated based on the adaptive mesh refinement (AMR) strategy. Please refer to \citet{Feng2011} and \citet{Yang2015} for the detailed description of the numerical methods. 

The initial conditions are set to be a state of hydrostatic equilibrium with zero initial velocity and potential magnetic field. Height profile of the temperature is defined by a hyperbolic tangent function to connect the chromospheric low temperature and coronal high temperature with a rapid transition. Potential magnetic field is constructed by adding an infinite series of line dipoles as described by \citet{Edmondson2010} to the uniform background field. The initial magnetic field has three poles rooted at the bottom boundary and an x-point geometry hanging in the numerical domain, thus forming a configuration favorable for occurrence of interchange magnetic reconnection. 

To produce guide field discontinuity and to drive the interchange MR between closed loop and open field, we introduce shear flows ($V_{\rm{z}}$) along the out-of-plane z-direction at the footpoints of closed loop. So $V_{\rm{z}}$ at the bottom boundary is set with 

  \begin{eqnarray}
V_{\rm{z}}=\left\{\begin{array}{cc}0, & \left|\left(x-x_{0}\right)\right| \geq r \\ 1.5 \sin \left(\pi\left(x-x_{0}\right) / r\right)\left(\left(x-x_{0}\right)^{2}-r^{2}\right)^{2}, & \left|\left(x-x_{0}\right)\right| \leq r\end{array}\right.
 \end{eqnarray}
where $x_{0}=27$~Mm for the center position of closed loop, and $r=15$~Mm for the half spanning distance of closed loop with shear flows. The shear flows at the footpoints of closed loop stretch the closed loop along the z-direction and lead to the production of guide field along that direction. On the contrary, the adjacent open field lines are not stretched and have no guide field since there are no shear flows at their footpoints. The distance and size of the closed loop under shearing motion is selected as typical values for closed loops of supergranular size \citep{Mariska1992}.

\section{Waves and Flows Launched from Interchange Reconnection in Simulation}
The stretched closed loop is also expanding upward to squash the neighboring open field lines, since the magnetic pressure of closed loop is enhanced with the extra contribution from newly produced guide field. A long and thin current sheet is formed between the squeezing closed loop and squeezed open field lines, and is subject to occurrence of tearing instability (see Figure 1a \& 1b). Plasmoids are generated between multiple x-points on the long current sheet, and ejected away by reconnection jet flows to bump into the open field region inducing secondary magnetic reconnection at the time of impact (see Figure 1c). The local thermal pressure at the impact place is further enhanced due to the second time of magnetic reconnection, and hence accelerates the jet flows more effectively and excites the compressive waves of larger amplitudes. Therefore, the jet flows and slow-mode waves in the newly reconnected flux tube on one side of the reconnection front lead to more remarkable difference of plasma parameters from its vicinity on the other side of the reconnection front.

These processes taking place at the border between closed loop and open fields give rise to the favorable conditions at the source region for the emission and bi-directional propagation of MHD waves: local enhancement of density and temperature ($\delta N\neq 0$, $\delta T\neq 0$), local discontinuity of guide field component ($\delta B_\perp \neq 0$). These initial disturbances facilitate the eigenmode decomposition of the disturbances into waves propagating in opposite directions.

The three components of fluctuating velocity vector ($\delta V_{\rm{x}}$, $\delta V_{\rm{y}}$, $\delta V_{\rm{z}}$) can basically be regarded as the orthogonal fluctuating velocities for co-propagating fast-mode, slow-mode, and Alfv\'en-mode waves, respectively. From Figure~1d, 1e, and 1f as well as the associated online Movie~1, we can observe the diverging propagation of $\delta V_{\rm{x}}$ at the fastest speed, the tube-guided upward propagation of $\delta V_{\rm{y}}$ and $\delta V_{\rm{z}}$ at the slowest and intermediate speeds, respectively. The enhancement of $V_{\rm{y}}$ component is contributed not only by the upward propagating slow-mode waves but also by the jet flows. For comparison, we conduct another case of simulation with the zero guide field due to the absence of shearing motion between the two legs of closed loop. We find that there are prominent differences between the two cases. For the zero-guide-field case, the compressive waves (fast- and slow-mode waves) exist with oscillating components in the simulation plane, while the incompressive waves (Alfv\'en waves) are absent without the oscillating components in the out-of-plane direction (see Movie 2).

Prominent enhancement of temperature also takes place in both the reconnected open funnel and reconnected closed loop (see the two bifurcated magnetic structures on two ends of the long thin current sheet in Figure~1c). The rise in temperature can be caused by three physical processes: (1) thermal conduction from the reconnected and heated source region; (2) convection of hot materials by the jet flows; (3) propagation of compression along with the slow-mode waves. 

The propagation signatures of the three types of MHD waves in the reconnected open funnel are also displayed as height-time diagrams in Figure 2. It can be seen that the height subinterval between 25~Mm and 30~Mm serves as the source region of wave emissions. The emission and propagation of waves occur intermittently with time and synchronously between the three types.  

\begin{figure}
\plotone{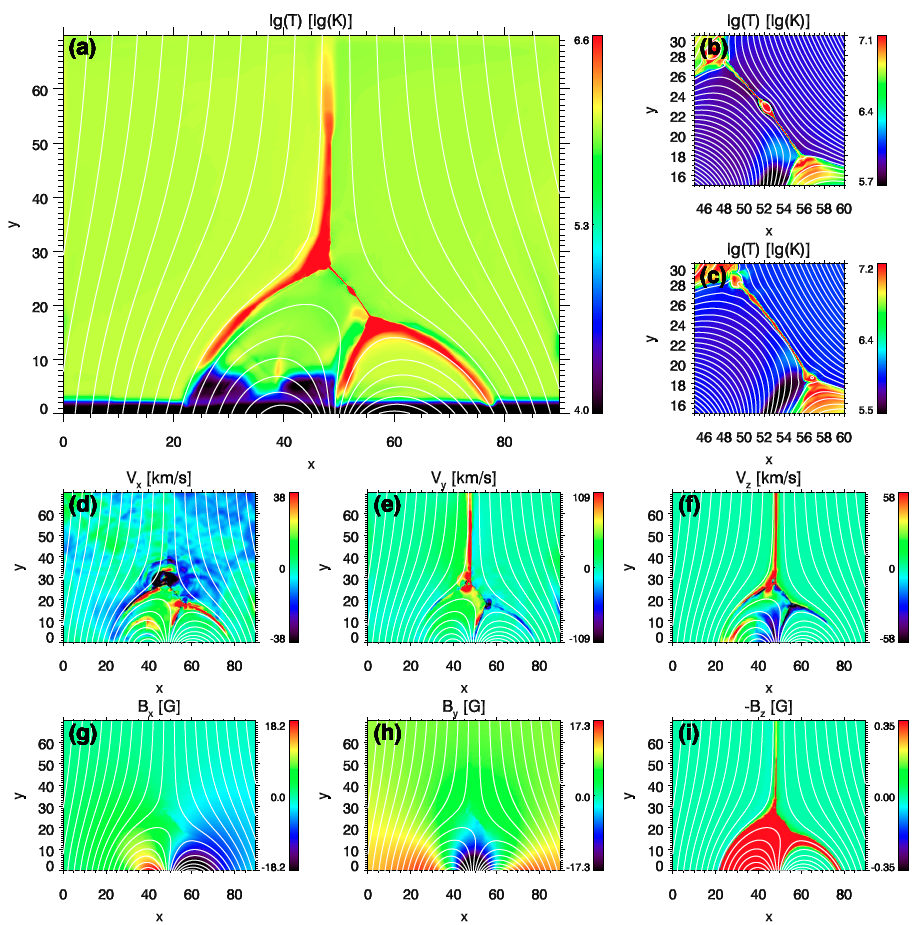}
\caption{Distribution of temperature (a-c), velocity (d-f), and magnetic field (g-i) in the source region of interchange magnetic reconnection with discontinuous guid field. The dynamic evolution of plasmoid, jet flows, fast-mode, slow-mode, and Alfv\'en-mode waves can be observed in the corresponding Movie~1.\label{fig:fig1}}
\end{figure}

\begin{figure}
\plotone{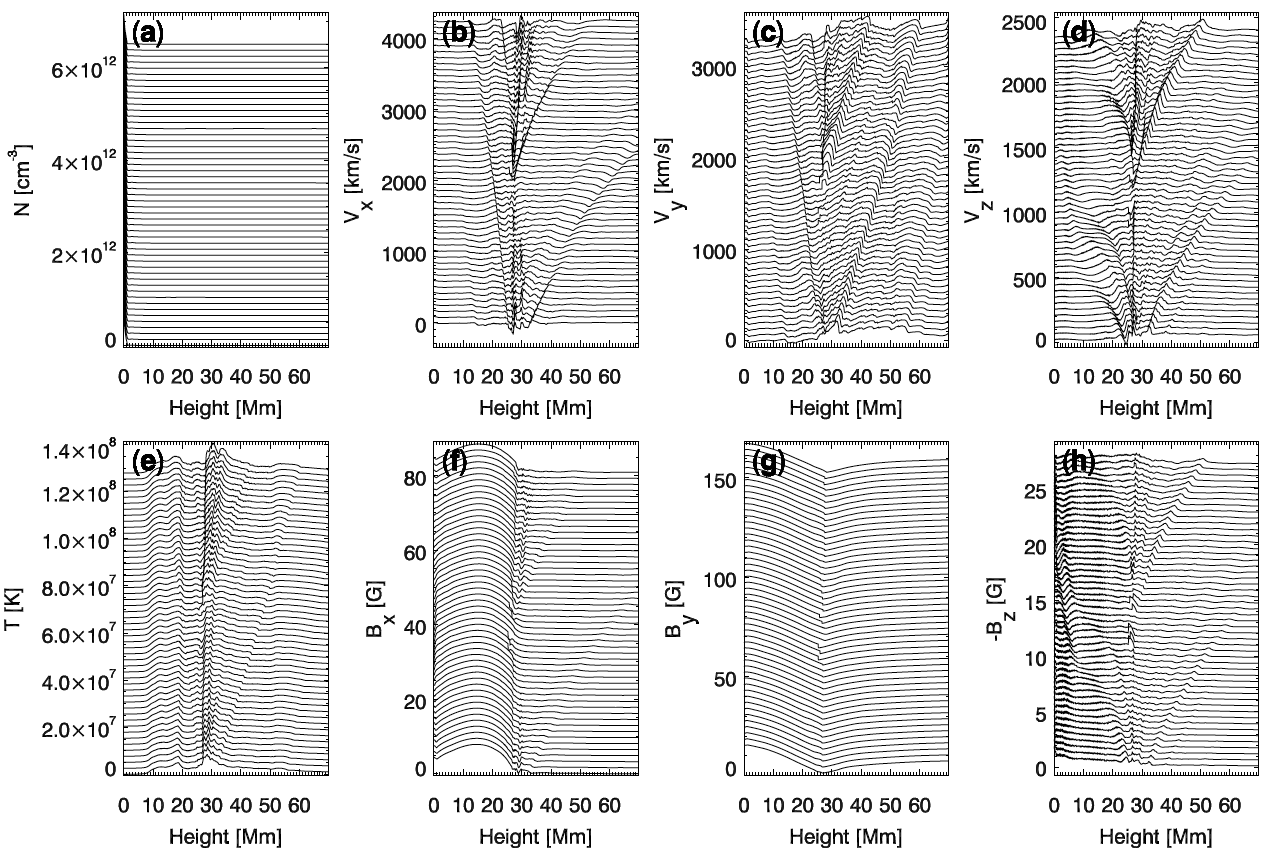}
\caption{Temporal variation of height profiles of variables along the newly reconnected open magnetic flux tube. $V_{\rm{x}}$, $V_{\rm{y}}$, and $V_{\rm{z}}$ basically represent the fluctuating velocities of fast-mode, slow-mode, and Alfv\'en-mode, respectively.\label{fig:fig2}}
\end{figure}

\section{Comparison between Simulation and Observation Results: Asymmetric Alfv\'enic Pulse, Temperature Enhancement, and Density Variation}

Since the spatial inhomogeneity in the inner heliosphere may stem from the inhomogeneous solar atmosphere, plotting and investigating the spatial profiles of variables across the reconnected open funnel at a certain height (see Figure~3) may help our understanding of the essence of inhomogeneous solar wind turbulence. In Figure 3a and 3b, we find asymmetric profile shapes of $\delta V_\perp$ (i.e., $\delta V_{\rm{z}}$) and $\delta B_\perp$ (i.e., $\delta B_{\rm{z}}$). The sharp and gradual transitions on the right and left sides of the peaks/dips represent respectively that the sampling point is entering from ambient un-reconnected region into newly-reconnected flux tube, and then exiting from the post-reconnected flux tube towards the ambient un-reconnected region. The occurrence of sharp transition in the spatial profile means that the sampler is encountered with a special structure, which we call it ``reconnection front'' as the border separating the reconnected and un-reconnected flux tubes. 

In the flux tube corresponding to reconnection front, one may observe some signals released from reconnection site to remotely sense what is going on at the reconnection site. For example, in the newly-reconnected flux tube (reconnection front), we can see the bumps of both density and temperature (See Figure 3c and 3d). In the newly-reconnected flux tube, the positive correlation between density and temperature bumps as well as the upward velocity together indicate the existence of slow-mode waves oscillating on one side unlike two-side oscillation of ordinary sinusoidal waves. While in the post-reconnected flux tubes (with some distance away from the reconnection front), there is a lowland in the density profile but a highland in the temperature profile (Figure 3c and 3d), which together with the intermittent upward flows suggest the formation of rarefaction region left behind the jet flows. We note that, both the ascending and descending phases of the Alfv\'en pulse are accompanied by an increase in temperature relative to the background, but not necessarily with an increase in density. The close relation between Alfv\'enic fluctuation and temperature enhancement is distinct from the weak relation between Alfv\'enic fluctuation and density variation. 

\begin{figure}
\begin{center}
\includegraphics[width=0.65\textwidth]{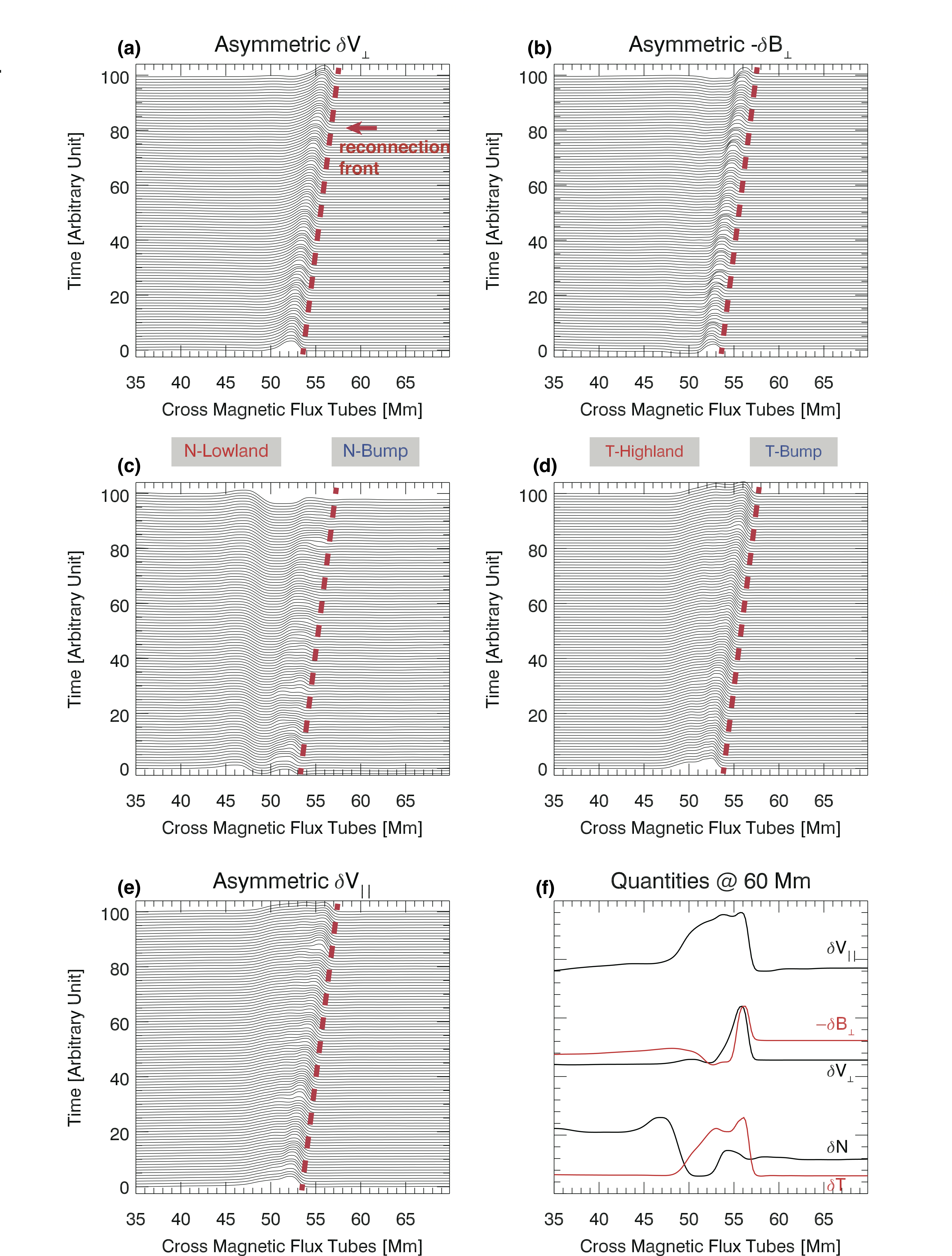}
\end{center}
\caption{Time stack plots of spatial profiles of variables ($V_\perp$ (a), $B_\perp$ (b), $N$ (c), $T$ (d), $V_\parallel$ (e), and all of them at a time (f)) across the un-reconnected, newly-reconnected, post-reconnected, and again un-reconnected flux tubes (from right to left). \label{fig:fig3}}
\end{figure}

The sampling direction in Figure 3 is in a uni-direction, supposed that all the three types of flux tubes (un-reconnected, newly-reconnected, and post-reconnected flux tubes) stay relatively static to each other in the x-z plane and wait for the sampler to pass through. In reality, the flux tubes may move randomly and abruptly relative to each other, causing that the sampler may meander through the flux tubes at different speeds. Here, we provide two examples in Figure 4 as the sampling results of two meandering passages. The first passage route is designed to enter into the newly-reconnected flux tube first and then return back to the un-reconnected flux tube at a slower speed. While the sampler in the second passage route first enter into the post-reconnected flux tube but no encounter of the newly-reconnected flux tube, and then leave for the un-reconnected flux tube again. 

According to our statistical studies of the PSP data, we find three types of magnetic deflection boundaries: (1) $\frac{T_{\rm{inside}}}{T_{\rm{outside}}}>1$ and $\frac{N_{\rm{inside}}}{N_{\rm{outside}}}>1$ for type-I; (2) $\frac{T_{\rm{inside}}}{T_{\rm{outside}}}>1$ and $\frac{N_{\rm{inside}}}{N_{\rm{outside}}}<1$ for type-II; (3) $\frac{T_{\rm{inside}}}{T_{\rm{outside}}}>1$ and $\frac{N_{\rm{inside}}}{N_{\rm{outside}}}\sim1$ for type-III, where the subscripts ``inside'' and ``outside" mean inside and outside the deflected magnetic structures, respectively. In Figure 4, we illustrate and compare the crossings of Alfv\'enic pulse/deflection boundaries between the simulation sampling results and the observational statistical results. The magnetic deflection/kink events are searched according to the two criteria: (1) $\theta_{\rm{BR}}$ is less than $120^\circ$ indicating the significant deviation from the background sunward magnetic field direction; (2) the duration is longer than 5 s. The tabular information of time intervals of these three types of events are provided as supplementary data files: ``list 1.txt'', ``list 2.txt '', and ``list 3.txt'', respectively.

In the first example, we can observe the enhancement of both temperature and density (Figure 4a) at the place of Alfv\'enic pulse (Figure 4b). Similar profiles can be found from PSP's of 25 events of type-I boundary crossing and are shown in Figure 4c and 4d. The earliest rise in temperature in the model sampling result is caused by the thermal conduction of heat flux, which is generated in the reconnection source region, along the edge of newly-reconnected flux tube. The following up rise in density corresponds to the upward propagation of slow-mode waves in the newly-reconnected flux tube. Such a refined difference between the rises of temperature and density in the modeling solar corona does not exist in the PSP's observations of the inner heliosphere.

In the second example, we see a rise in temperature but a decline in density as associated with asymmetric Alfv\'enic fluctuations (Figure 4e and 4f). As a consistent comparison, we present the PSP's 29 events of type-II boundary crossing in Figure 4g and 4h, showing that temperature jumps up while density falls down at the sharp transition (ascending) phase of Alfv\'enic pulse. We also find 22 events of type-III boundary crossing events.

\begin{figure}
\plotone{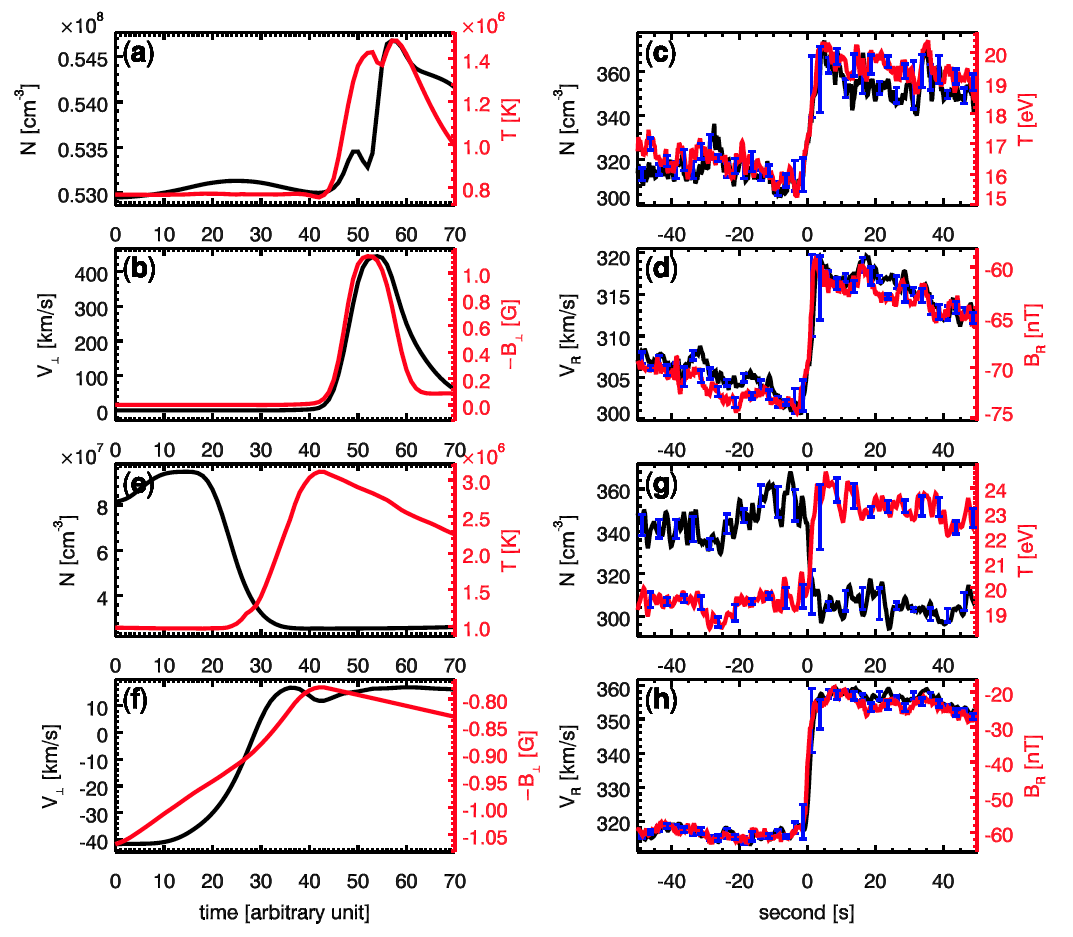}
\caption{Comparison of two types of Alfv\'enic pulses between simulation results (left) and PSP observations (right). (a \& b) Sampled profiles of $N$, $T$, $V_\perp$, and $B_\perp$ for type-I. (c \& d) Observed $N$, $T$, $V_{\rm{R}}$, and $B_{\rm{R}}$ from PSP for type-I, a superposed-epoch analysis result of 25 events. (e \& f) Sampled profiles of $N$, $T$, $V_\perp$, and $B_\perp$ for type-II. (g \& h) Observed $N$, $T$, $V_{\rm{R}}$, and $B_{\rm{R}}$ from PSP for type-II, a superposed-epoch analysis result of 29 events. \label{fig:fig4}}
\end{figure}

The features in common between these two types lie in the temperature increase and asymmetric Alfv\'enic pulse fluctuations due to the encounter of reconnected open flux tubes. The difference between these two types lies in different density variations due to the different evolutionary phases of reconnected flux tubes: newly-reconnected phase for the first type, and post-reconnected phase for the second type.  According to the WKB approximation of Alfv\'en wave propagation, the velocity disturbance may surpass the local Alfv\'en speed (i.e., becoming super-Alfv\'enic fluctuation and likely switchback of magnetic field line) at the altitudes of [1, 10]~Rs and [10, 50]~Rs when $\delta V$ is on the order of 100 km/s and 10 km/s at the coronal source region, respectively. The quantitative relation of variables between the simulation at the source region and the PSP observation in the inner heliosphere is regarded as a challenging issue since the reflection, dissipation, conversion of wave modes along the propagation cannot be easily tackled.

\section{Discussion}

The time-varying geometries of magnetic field lines involving interchange reconnection are displayed in Figure 5. There exist various types of upward energy fluxes in terms of heat flux by thermal conduction, wave energy fluxes of Alfv\'en waves and slow-mode waves, as well as bulk kinetic energy flux of jet flows. The altitude evolution of the waves and jet flows is a crucial issue to connect the origin of waves and flows in the solar atmosphere with the in-situ measurements of their development in the heliosphere. To extend the numerical model to higher altitude beyond 1~$R_{\rm{s}}$, we need to adopt the spherical coordinate system, in which nonlinear evolution of waves in the (super-)radial expanding flux tubes can be investigated more naturally. The stronger $\delta V/V_{\rm{A}}$ at the source region, e.g., Alfv\'en pulse generated via interchange reconnection with discontinuous guide field, is likely to exceed 1 after nonlinear evolution in the inner heliosphere, indicating the occurrence of super-Alfv\'enic turbulence as well as the switchback magnetic field configuration. The formation of a magnetic switchback geometry during the radial nonlinear evolution of Alfv\'enic turbulence is preliminarily reproduced through the approach of expanding-box MHD simulation \citep{Squire2020}.

\section{Conclusion}

In this work, we proposed and conducted a model involving interchange reconnection between open and closed magnetic field lines with discontinuous guide fields, which are produced self-consistently due to the shearing motions at the two footpoints of closed loop. The reconnection with guide field discontinuity of this work can be generalized to be counter-helicity reconnection when considered in a 3D scenario.

We find that the three types of MHD waves can be excited quasi-periodically at both ends of a long and thin current sheet, which intermittently triggers multiple x-point magnetic reconnection and ejects plasmoids. We note that the slow-mode waves only propagate in the newly-reconnected flux tube, while the Alfv\'en waves exist in both newly-reconnected and post-reconnected flux tubes. This means that emission of Alfv\'en waves in the reconnected flux tube exists for a longer time than that of slow-mode waves. The longer time of Alfv\'en wave emission is due to that it takes longer time for the Alfv\'enic transverse velocity to complete the transverse displacement and hence to restore the reconnected geometry back to a normal geometry of open field line. In contrast, the emission of slow-mode waves just last for a relatively shorter time until the propagation away and disappearance of thermal pressure enhancement originally induced by the impact of ejected plasmoids. We also note that, as the jet flow moves upward, the post-reconnected region left behind forms a rarefaction region characterized with lower density and higher temperature than ambient un-reconnected background plasmas. 

The comparison between the sampling of numerical results and the observational results helps us to infer what type of structure PSP is encountered when it measures the Alfv\'enic velocity spikes or magnetic field switchbacks. The encounter of newly-reconnected flux tubes (``reconnection front'') by PSP may be responsible for its observed asymmetric Alfv\'enic pulses as well as its observed increases in temperature and density. It seems plausible that PSP passes through post-reconnected flux tubes and observes the asymmetric Alfv\'enic pulses in association with an increase in temperature but a decrease in density. The in-situ measurements of asymmetric Alfv\'enic pulses from PSP hence cast new light to trace the evolution of interchange reconnection in the solar wind source region.

\begin{figure}
\plotone{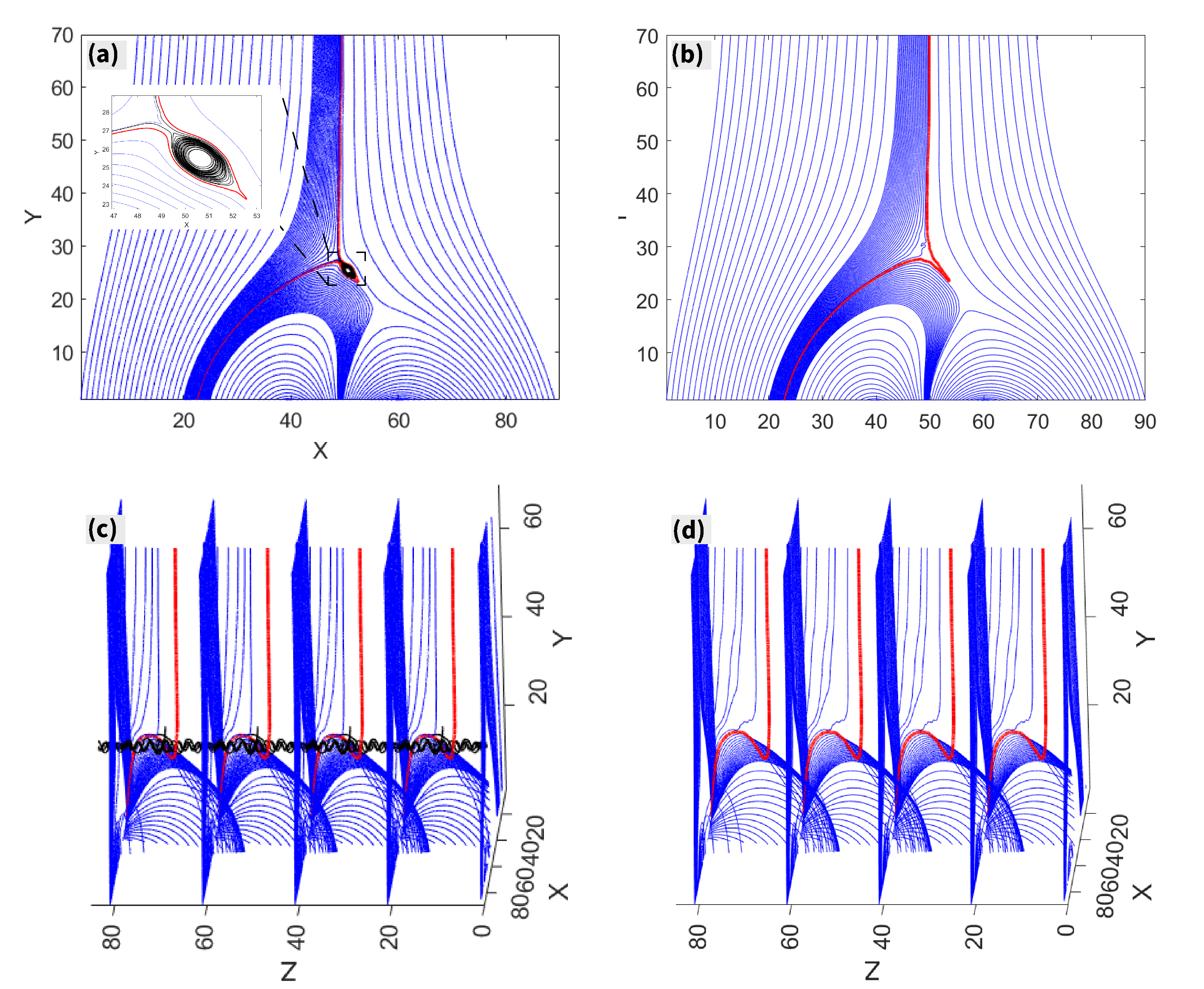}
\caption{Changes of magnetic field geometries associated with the Alfv\'enic pulse release events observed from different viewing angles. Magnetic flux ropes (plasmoids) (black lines) as formed in the long current sheet impact into the open field lines and disappear after the collisions. \label{fig:fig5}}
\end{figure}

\section{Acknowledgements}
This work is supported by NSFC under contracts 41874200, 41774157, 41974171, and 41804164, as well as by CNSA under contracts No. D020301 and D020302. L.P.Y is the co-corresponding author of this work. The authors acknowledge the PSP team for making the data available on SPDF (https://cdaweb.sci.gsfc.nasa.gov/index.html/).


\bibliography{references}
\bibliographystyle{aasjournal}

\end{document}